\documentclass[twocolumn]{aastex62}

\usepackage{amsmath}

\received{\today}
\revised{...}
\accepted{...}

\submitjournal{ApJ}

\shorttitle{Modulation of the radio emission from the solar corona by a fast wave}
\shortauthors{Kolotkov et al.}

\begin{document}

\title{{The origin of the} modulation of the radio emission from the solar corona by a fast magnetoacoustic wave}

\correspondingauthor{Dmitrii Y. Kolotkov}
\email{D.Kolotkov@warwick.ac.uk, D.Kolotkov.1@warwick.ac.uk}

\author{Dmitrii Y. Kolotkov}
\affil{Centre for Fusion, Space and Astrophysics, Department of Physics, University of Warwick, CV4 7AL, UK}

\author{Valery M. Nakariakov}
\affiliation{Centre for Fusion, Space and Astrophysics, Department of Physics, University of Warwick, CV4 7AL, UK}
\affiliation{School of Space Research, Kyung Hee University, Yongin, 446-701, Gyeonggi, Korea}

\author{Eduard P. Kontar}
\affiliation{School of Physics and Astronomy, University of Glasgow, Glasgow G12 8QQ, UK}

\begin{abstract}

Observational detection of quasi-periodic drifting fine structures in a type III radio burst associated with a solar flare SOL2015-04-16T11:22, with {Low Frequency Array}, is presented.
{Although similar modulations of the type III emission have been observed before and were associated with the plasma density fluctuations, the origin of those fluctuations was unknown.}
Analysis of the striae of the intensity variation in the dynamic spectrum allowed us to reveal two quasi-oscillatory components. The shorter component has the apparent wavelength of {$\sim2$\,Mm}, phase speed of $\sim657$\,km\,s$^{-1}$, which gives the oscillation period of $\sim3$\,s, and the relative amplitude of $\sim0.35$\%. The longer component has the wavelength of {$\sim12$\,Mm}, and relative amplitude of $\sim5.1$\%. The short frequency range of the detection does not allow us to estimate its phase speed. {However}, the properties of the shorter oscillatory component allowed us to interpret it as a fast magnetoacoustic wave guided by a plasma non-uniformity along the magnetic field outwards from the Sun. The assumption that the intensity of the radio emission is proportional to the amount of plasma in the emitting volume allowed us to show that the superposition of the plasma density modulation by a fast wave and a longer-wavelength oscillation of an unspecified nature could readily reproduce the fine structure of the observed dynamic spectrum. The observed parameters of the fast wave give the absolute value of the magnetic field in the emitting plasma of $\sim1.1$\,G which is consistent with the radial magnetic field model.

\end{abstract}


\section{Introduction} \label{sec:intro}
Fast wave trains are the localised in space and time quasi-periodic disturbances of plasma density, formed by the dispersive evolution of fast magnetohydrodynamic (MHD) waves excited by a broadband driver, and propagating along a plasma non-uniformity which acts as a waveguide \citep{1983Natur.305..688R,1984ApJ...279..857R, 2004MNRAS.349..705N}. In the corona, the typical speed of the fast wave trains is about the Alfv\'en speed, i.e. 1000--4000\,km\,s$^{-1}$, and the dominant period is prescribed by the Alfv\'en transit time across the waveguiding plasma non-uniformity, typically from a few seconds to a few minutes. Comparison of observational properties of the fast wave trains to predictions of MHD theory could allow for a seismological inversion of the local plasma parameters along the path of the wave train propagation \citep[see][for recent comprehensive reviews ]{2016SSRv..200...75N, 2018SSRv..214...45M}. 

Fast wave trains are directly observed in the lower corona with imaging telescopes.
Characteristic signatures of fast wave trains have been found in rapidly-propagating, at the apparent phase speed of 2100\,km\,s$^{-1}$, quasi-periodic, with the mean period of about 6~s, perturbations of the white-light intensity in the lower corona (around $1.1R_\odot$) during a total solar eclipse by \citet{2001MNRAS.326..428W, 2004MNRAS.349..705N}. More recently, rapidly-propagating quasi-periodic disturbances of the extreme ultraviolet (EUV) emission have been detected with the Atmospheric Imaging Assembly onboard the Solar Dynamics Observatory (SDO/AIA). The first unambiguous detections of such propagating quasi-periodic disturbances of the 171\,\AA\ and 193\,\AA\ emissions were reported by \citet{2011ApJ...736L..13L,2012ApJ...753...52L}. The disturbances were associated with the fast magnetoacoustic wave trains propagating along coronal funnel-like structures, with a plane of the sky speed of 650--2000\,km\,s$^{-1}$ and the period of 2--3 min. Similar observational studies were carried out by e.g. \citet[][the projected phase speed of $\sim834$\,km\,s$^{-1}$ and mean period of 25--83\,s]{2012ApJ...753...53S}, \citet[][600--735\,km\,s$^{-1}$ and 38--58\,s]{2013A&A...554A.144Y}, \citet[][$\sim1000$\,km\,s$^{-1}$ and 60--80\,s]{2014A&A...569A..12N}, and in a more recent work by \citet[][875--1485\,km\,s$^{-1}$ and 75--160\,s]{2018ApJ...853....1S}. 

In addition to the direct observations in EUV, characteristic signatures of dispersively evolving fast wave trains have been indirectly detected in the radio emission. Characteristic signatures of fast wave trains in type IV radio bursts were for the first time recognised by \citet{2009ApJ...697L.108M, 2011SoPh..273..393M}, in the 1.1--4.5\,GHz emission. The detected periods are 71--81\,s, similar to those seen in the direct EUV observations, and also 0.5--1.9\,s. \citet{2016A&A...594A..96G} detected a train of quasi-periodic radio sparks at 42--83\,MHz, which had the repetition rate of about 1.8~ min, similar to the periodicity detected in EUV emission disturbances. The observational frequencies corresponded to the heights from $1.3R_\odot$ to $1.6R_\odot$, according to the Newkirk density model of the corona \citep{1961ApJ...133..983N}, and time delay between the EUV and radio observations was consistent with the propagation at the Alfv\'en speed. \citet{2017ApJ...844..149K} showed evidence of fast wave trains in type IV and type III radio bursts at 245\,MHz and 610\,MHz, with periods of about 70--140\,s, which were found to be similar to the periodicity detected in quasi-periodic waves of the EUV emission at 171\,\AA\ and 193\,\AA.  

Fast wave trains are a promising seismological probe of the {middle and} upper corona, above the field-of-view of EUV images and spectrographs. This region attracts specific interest in the context of the solar wind acceleration. At those heights wave processes could be studied by coronagraphs in the white light, and with low-frequency radio instruments. For example, 9-min quasi-periodic variations in the white-light polarised brightness, detected by \citet{1997ApJ...491L.111O} in a coronal hole at heliocentric distances around $2R_\odot$ could be associated with a fast wave train.
{However, only low frequency radio observations can provide diagnostics of the corona at the sub-second scales at the coronagraph height. Moreover, unlike the line-of-sight integrated white light coronagraph observations, the radio emission highlights the local perturbations of plasma along the magnetic field, where it originates.}

In this paper, we present the observational detection of signatures of a quasi-periodic fast magnetoacoustic wave train propagating in the corona, and modulating a metric type III radio burst. These signatures are detected in the 35--39\,MHz frequency band, corresponding to the range of heliocentric distances of 1.6--1.7\,$R_\odot$ (using the Newkirk model), which makes this observation the highest detection of fast wave trains in the solar atmosphere in the radio band.
{Similar quasi-periodic modulations of the type III radio emission, highly likely related to the plasma density fluctuations, were detected in earlier observations \citep[see e.g. the original work by][]{1972A&A....20...55D}. However, the association of those density fluctuations with a specific MHD wave, propagating upwards through the stratified plasma of the solar corona, is conducted for the first time in the present paper.}

The observational data and its analysis providing the estimation of the propagating wave speed, wavelength, and period are described in Sec.~\ref{sec:obs}. A quantitative model for the observed modulation of the radio flux, based on its redistribution on the plasma density inhomogeneities produced by the wave, is presented in Sec.~\ref{sec:mech}. Furthermore, we estimate the wave energy flux and the ambient magnetic field strength and compare them with the local radiative energy losses and the radial model of the magnetic field in Secs.~\ref{sec:est} and \ref{sec:est_mf}, respectively. The summary of the obtained results and our conclusions are given in Sec.~\ref{sec:conc}.

\section{Analysis of observations} \label{sec:obs}

\begin{figure*}
	\begin{center}
		\includegraphics[width=0.35 \linewidth]{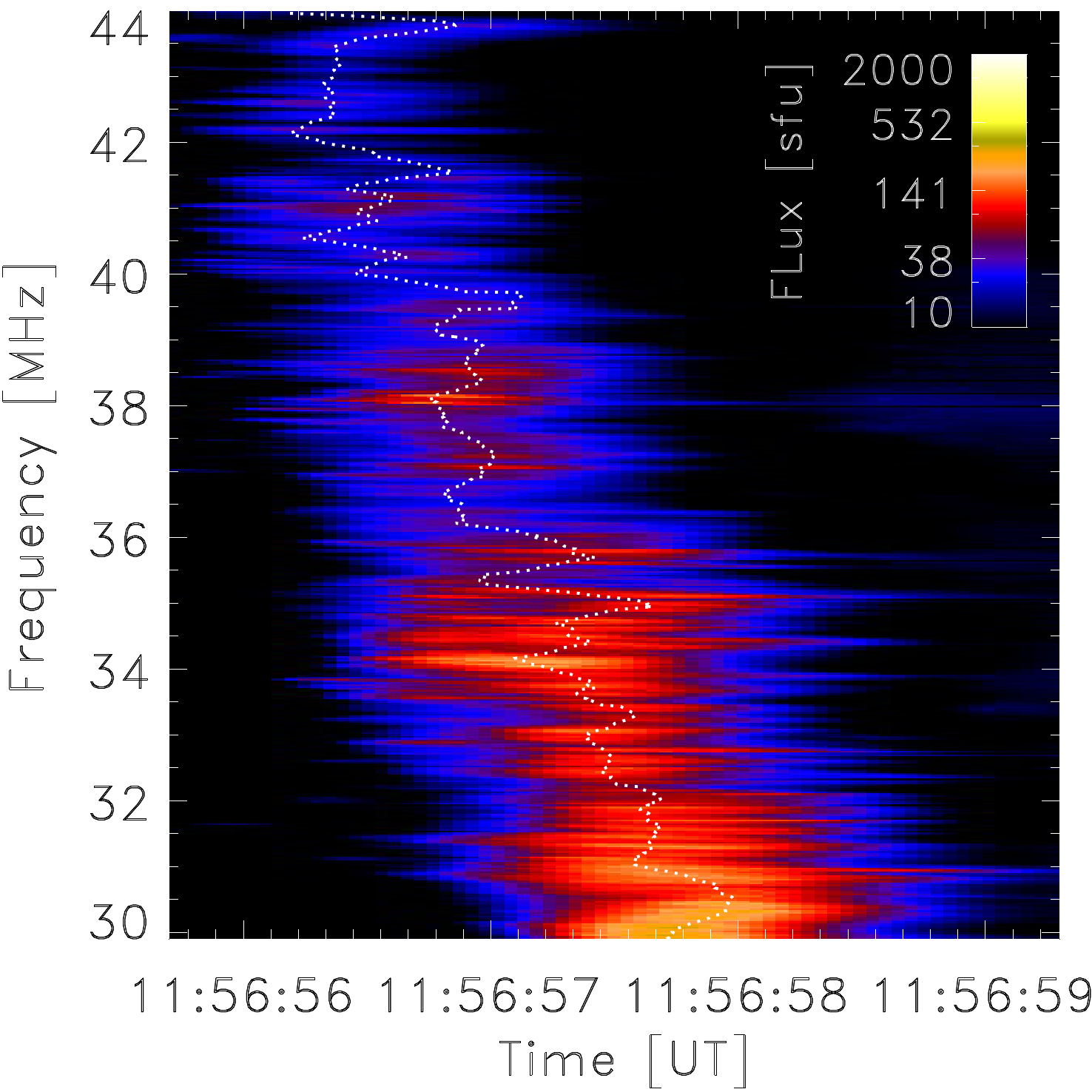}
		\includegraphics[width=0.35 \linewidth]{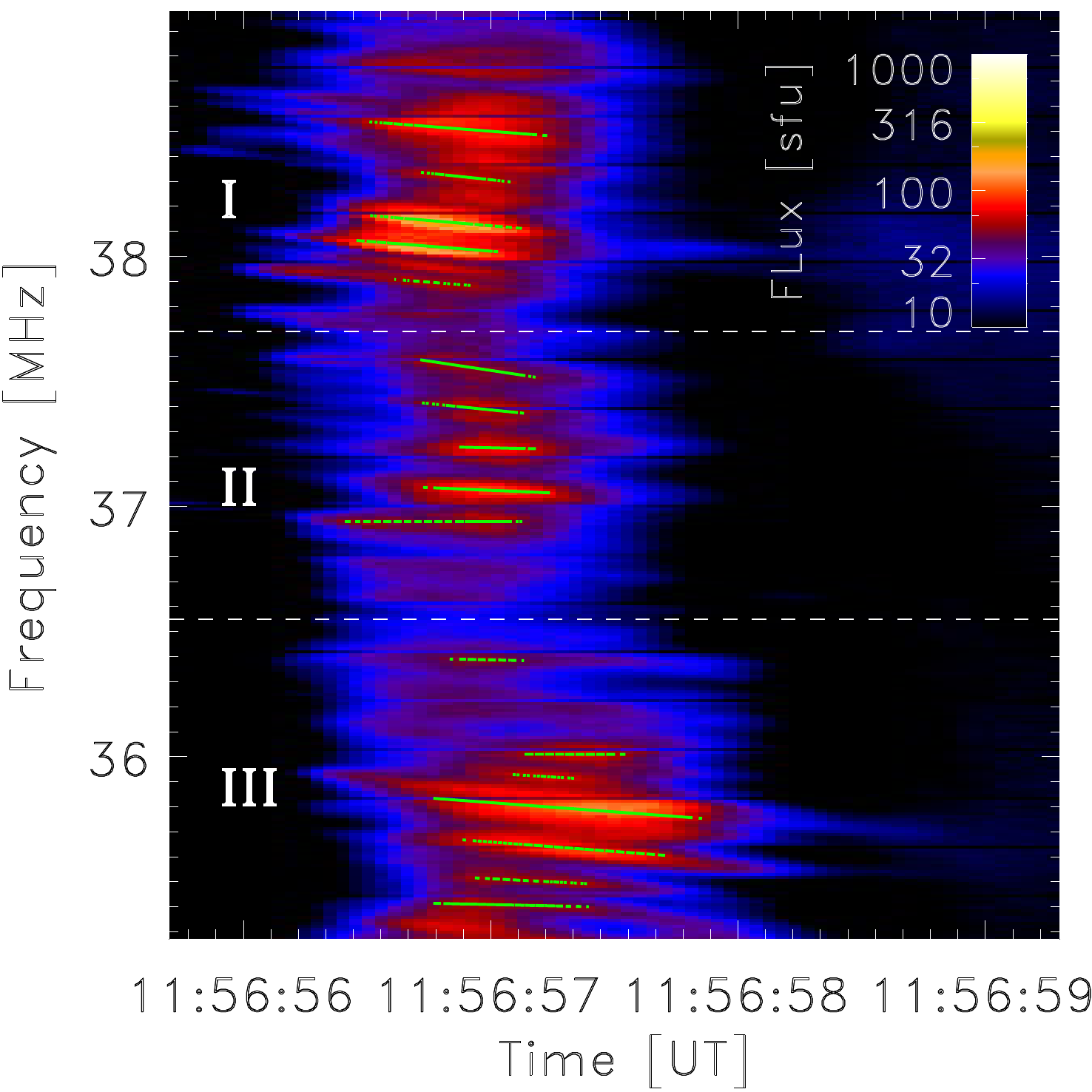}
	\end{center}
	\caption{A 3-s interval of the dynamic spectrum, that is a Sun-integrated radio flux in the frequency--time plane, of a type III solar radio burst occurred on 2015 April 16, and observed by the \emph{Low Frequency Array} (LOFAR). The left-hand and right-hand panels show the burst in the 30--44~MHz and 35--39~MHz frequency intervals, respectively. The white dotted line in the left-hand panel shows the \lq\lq spine\rq\rq\ of the burst, i.e. the instants of time of a maximum radio flux at each observational frequency. The straight green lines in the right-hand panel show fitting of the observed striae by a linear function. {The regions of apparent clustering of the striae into three distinct groups are indicated as \lq\lq I\rq\rq, \lq\lq II\rq\rq, and \lq\lq III\rq\rq\ and separated by the horizontal dashed lines in the right-hand panel.}
		\label{fig:spectra}}
\end{figure*}

\begin{figure}
	\begin{center}
		\includegraphics[width=\linewidth]{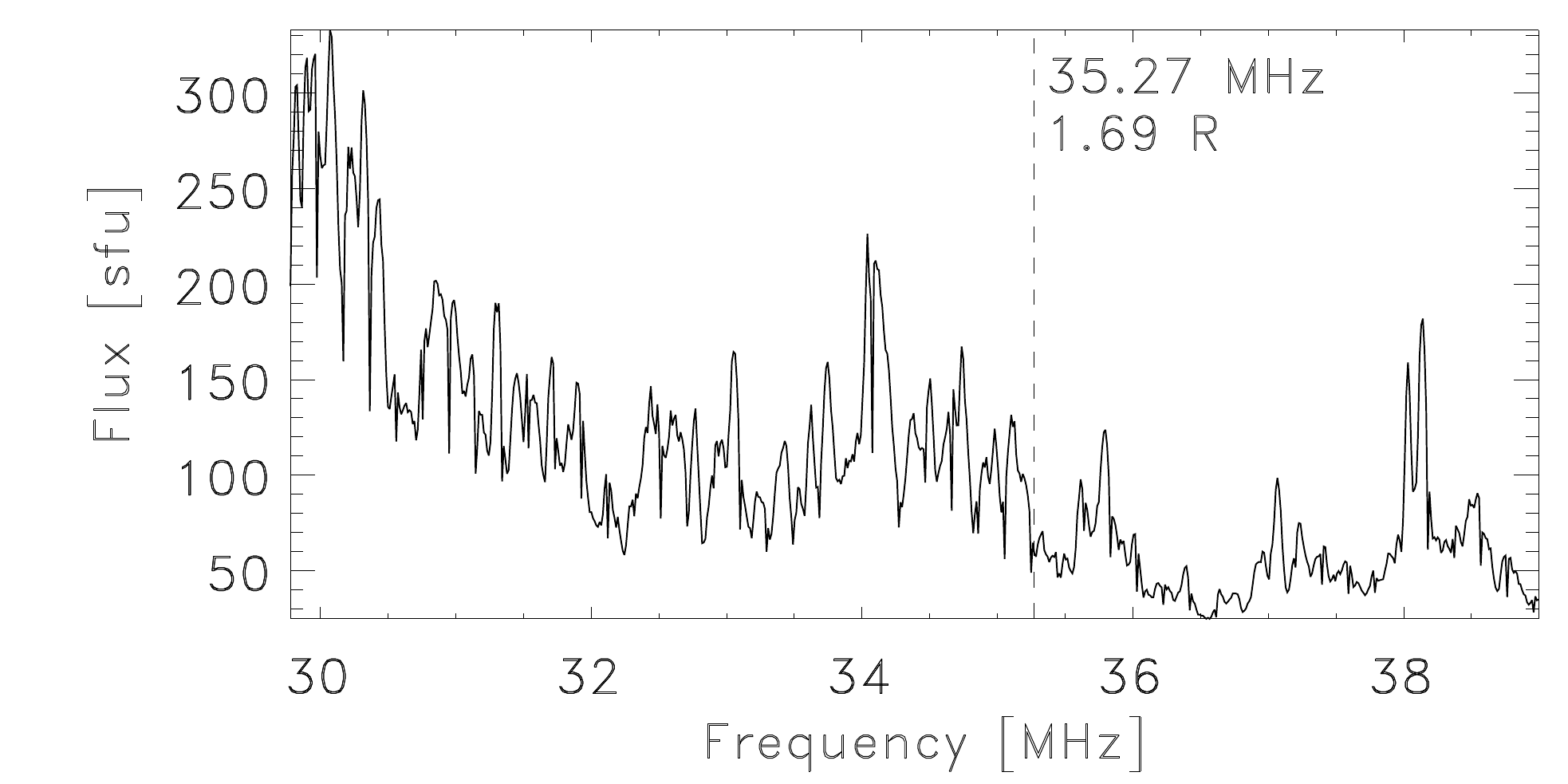}
		\includegraphics[width=\linewidth]{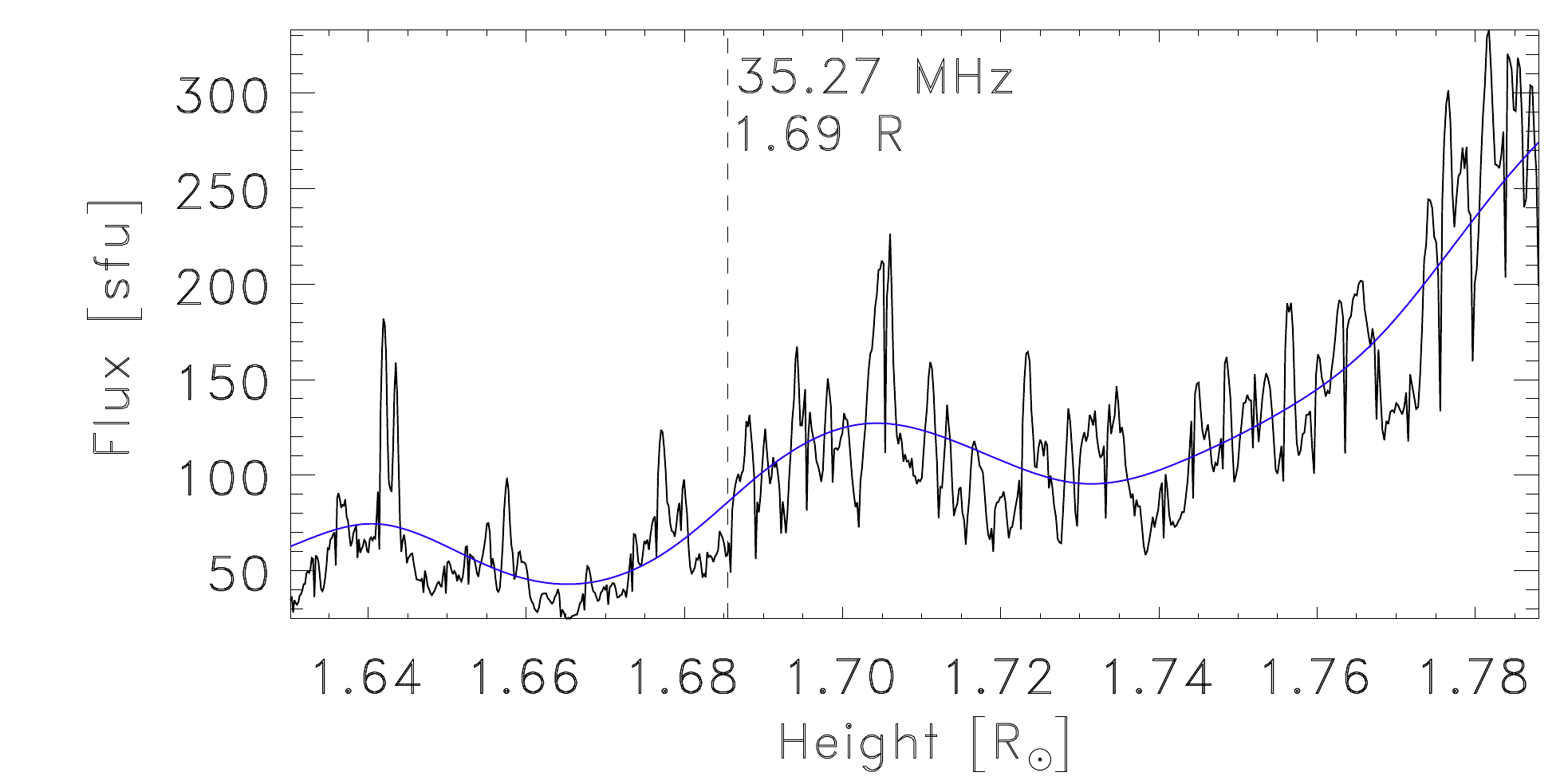}
	\end{center}
	\caption{The flux along the spine of the type III radio burst (see Fig.~\ref{fig:spectra}) as a function of frequency (top) and distance from the centre of the Sun (bottom). The vertical dashed lines in both panels show the frequency/height below/above which the quasi-periodic behaviour is poorly detected (see Figs.~\ref{fig:wavelet_1} and \ref{fig:wavelet_2}, and the corresponding description in Sec.~\ref{sec:obs}). The blue solid line in the bottom panel shows an overall trend of the observational signal, obtained using the empirical mode decomposition technique.
		\label{fig:flux}}
\end{figure}

The analysed type III solar radio burst occurred on 2015 April 16, after a C2.3-class flare in the active region NOAA 12321, peaked at 11:22:00\,UT and followed by a number of smaller-scale energy releases at the time of the burst. The burst was observed by the \emph{Low Frequency Array} (LOFAR) instrument \citep{2013A&A...556A...2V}, one of the largest ground-based decameter arrays that permits imaging and provides a high time resolution. The radio burst was produced by a beam of electrons accelerated up to $10^{10}$~cm\,s$^{-1}$ which is about 30\% of the speed of light, and reached the peak flux density of 100--200~sfu in the observed frequency interval between 32 and 40~MHz. Detailed quantitative analysis of the frequency and imaging information available for this event, including the size, location, and motion of the radio source, was recently performed by \cite{2017NatCo...8.1515K}.


The dynamic spectrum of the analysed burst is given in Fig.~\ref{fig:spectra}. It shows a radio flux in the frequency--time plane, with a rapid decrease of frequency with time, which is rather typical for the events of this type.  The dynamic spectrum has a fine structure in a form of quasi-periodic fluctuations appearing around 11:56:56--11:56:59\,UT, whose revealing became possible due to an exclusively high spectral resolution of LOFAR operating in 12~kHz-wide frequency channels. This fine structuring is represented by a number of nearly horizontal (in comparison with the slope of the burst in the frequency--time plane) striae, narrow lanes of the enhanced radio emission, with the modulation depth of about 25\% and the population density growing with decreasing frequency, and eventually overlapping around 30--31~MHz. These preliminary observational facts may apparently indicate that an electron beam producing the radio burst at the electron plasma frequency, propagates upwards through the stratified plasma of the solar corona, modified by a travelling wave. The phase speed of the modulating wave is much lower than that of the beam.
{This suggestion nicely extends the recent work by \cite{2018ApJ...856...73C}, who performed a statistical analysis of the radio flux fluctuations observed in this event. However, their origin was not established.}
A more detailed consideration supporting this suggestion is presented in the following analysis.

Figure~\ref{fig:flux} shows the variation of the radio flux with the frequency, calculated along the spine of the burst as shown in Fig.~\ref{fig:spectra}. Using the dependence of the electron plasma frequency upon the density, 
\begin{equation}
\label{eq:freq}
f_\mathrm{pe}[\mathrm{MHz}]\approx 9\times10^{-3} \sqrt{n[\mathrm{cm}^{-3}]},
\end{equation}
and assuming the Newkirk model of the solar corona \citep{1961ApJ...133..983N}, so that 
\begin{equation}
\label{eq:newkirk}
n(r)=n_0\,10^{4.32/r},
\end{equation}
where $n_0=4.2\times10^4$\,cm$^{-3}$ and $r$ is the heliocentric distance measured in units of the solar radius, $R_\odot$, one can obtain the dependence of the observed radio flux upon the height $r$, which is also shown in Fig.~\ref{fig:flux}. The intensity of the signal is stronger towards lower frequencies (grows with height), which is likely related to the growth of Langmuir waves and their conversion into the emission. Further away from the Sun, it is easier for the non-thermal electrons to generate stronger emission, 
because of the growing ratio of the densities of the non-thermal {electrons} to background plasma electrons. 

\begin{figure}
	\begin{center}
		\includegraphics[width=\linewidth]{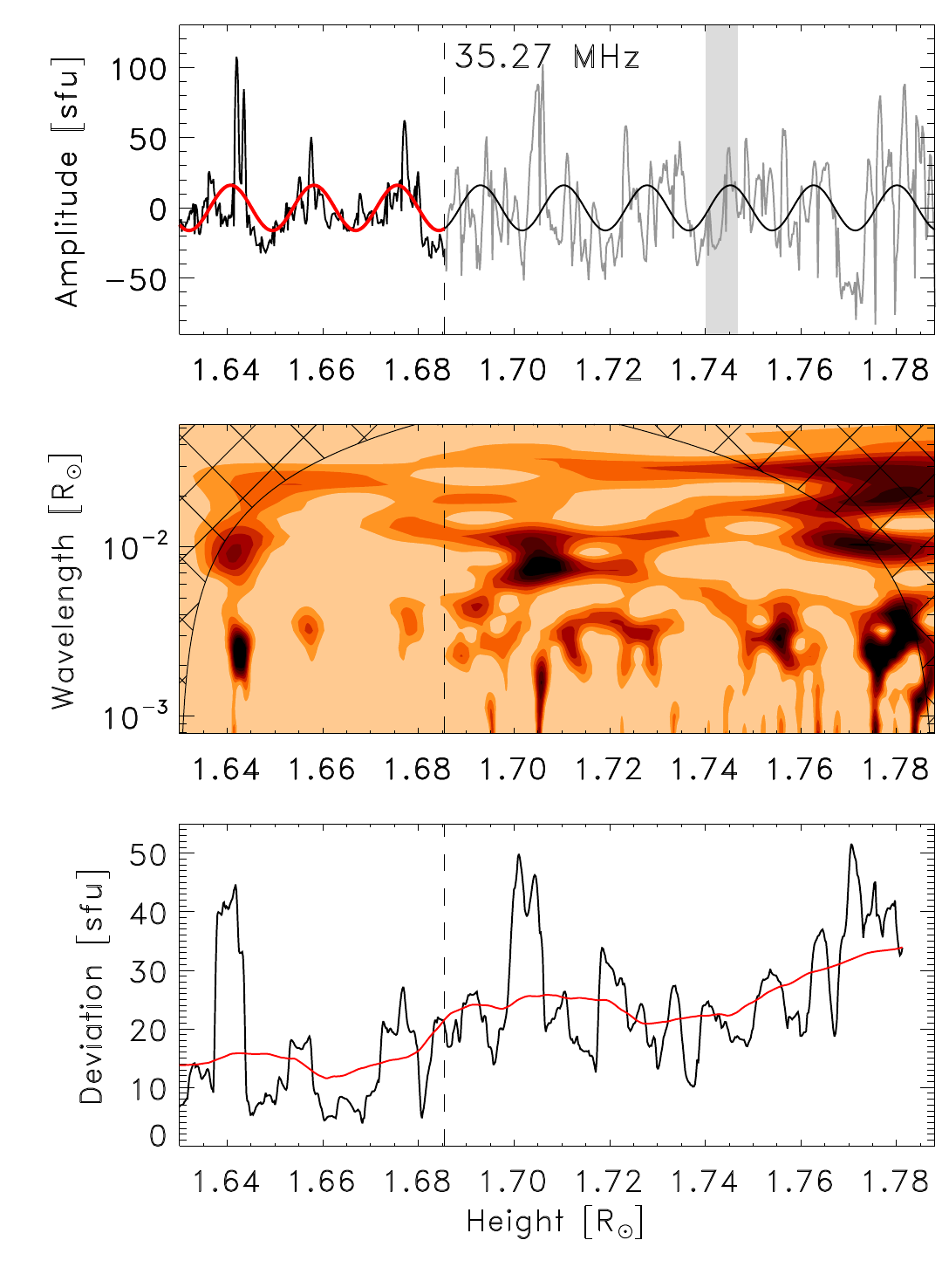}
	\end{center}
	\caption{
		Top: the flux along the spine of the type III burst (see Figs.~\ref{fig:spectra} and \ref{fig:flux}) as a function of distance from the centre of the Sun, with the overall trend subtracted. Red and black sinusoidal lines show its fitting by a harmonic function. The vertical dashed line in all panels are similar to those in Fig.~\ref{fig:flux}. The shaded rectangle illustrates a running window of about 0.007\,$R_\odot$ width (shown at the random location), within which the standard deviation of the detrended observational signal was calculated (see the bottom panel).
		Middle: the Morlet wavelet spectrum of the detrended observational signal shown in the top panel.
		Bottom: the standard deviation (black) of the detrended observational signal, calculated within a running window shown in the top panel; the red line shows the standard deviation smoothed with a boxcar average of about 0.05\,$R_\odot$ width.
	}
	\label{fig:wavelet_1}
\end{figure}

Subtracting the overall trend from the observed dependence of the radio flux upon the height, determined with the use of the empirical mode decomposition technique \citep[see][for details]{1998RSPSA.454..903E,2008RvGeo..46.2006H}, we calculate the Morlet wavelet spectrum of the detrended signal, shown in Fig.~\ref{fig:wavelet_1}. A visual inspection of the obtained spectrum shows the presence of a systematic behaviour in the observational signal at heights up to approximately $1.69\,R_\odot$ (about 35.27~MHz), above which its spectral energy behaves rather stochastically and there is no pronounced periodic components. The latter is also confirmed by the behaviour of the standard deviation calculated within a narrow running window (see Fig.~\ref{fig:wavelet_1}), whose mean value increases by a factor of 2 when passing through $1.69\,R_\odot$. Such a sudden change in the observed behaviour can be attributed to an increased impact of noise or development of local turbulent processes \citep[see e.g.][]{2017SoPh..292..155M,2018ApJ...856...73C}, which are out of the scope of this paper.

\begin{figure}
	\begin{center}

		\includegraphics[width=\linewidth]{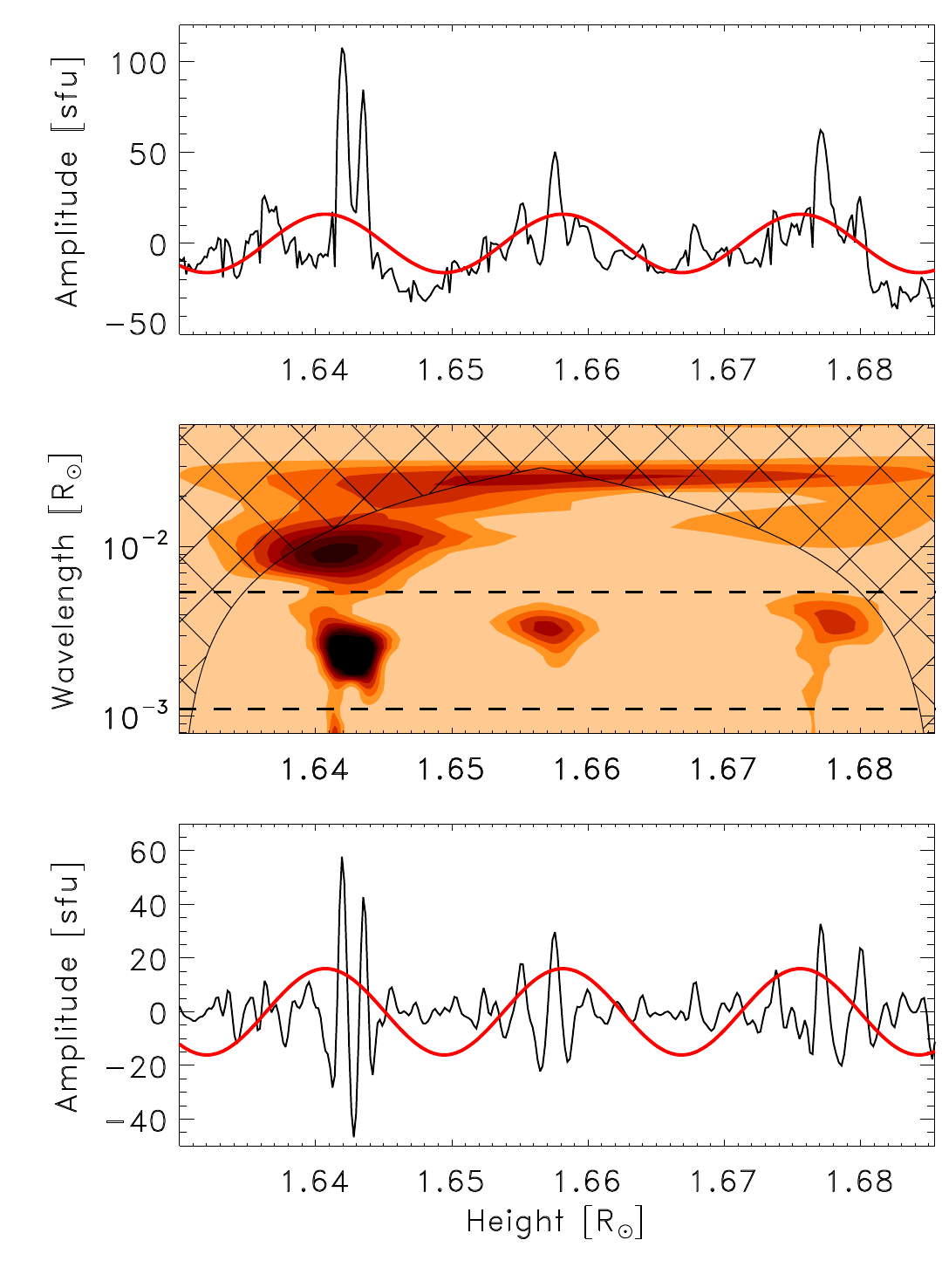}
	\end{center}
	\caption{The top and middle panels are similar to those shown in Fig.~\ref{fig:wavelet_1}, but for the region of interest from $1.63\,R_\odot$ to $1.69\,R_\odot$, where the quasi-periodic behaviour is the most pronounced.
	The bottom panel shows the periodic component with the wavelength of $(3\pm0.8)\times10^{-3}$\,$R_\odot$ (the black line), filtered out from the wavelet spectrum (the width of the narrow-band filter is shown by the horizontal dashed lines in the middle panel), and the best-fitting curve of the other oscillatory component, with the wavelength of $(17.4\pm0.2)\times10^{-3}\,{R}_\odot$ (the red line) obtained from the fitting of the detrended observational signal by a harmonic function.
	}
\label{fig:wavelet_2}
\end{figure}

In the following analysis, we therefore focus on the quasi-periodic behaviour of the observed radio flux at the approximate heights from $1.63\,R_\odot$ to $1.69\,R_\odot$, whose Morlet wavelet spectrum is shown in Fig.~\ref{fig:wavelet_2}. This spectrum shows two well-defined periodic components, one of which has an approximate wavelength of $(3\pm0.8)\times10^{-3}$\,$R_\odot$, and is associated with the striation seen in the dynamic spectrum in Fig.~\ref{fig:spectra}. The other detected periodicity is of a substantially longer wavelength, corresponding to the effect of clustering of striae in three distinct groups, {which was not detected in the earlier observations.} The longer-wavelength periodicity is situated near the boundary of the cone of influence, where edge effects become important. The latter makes an accurate quantitative estimation of its wavelength subjective, and therefore requires a separate analysis. To cope with this, we fit the observed signal by a harmonic function with the wavelength of $(17.4\pm0.2)\times10^{-3}\,{R}_\odot$, where the corresponding uncertainties are estimated within 95\% credible intervals using Bayesian inference and Markov chain Monte Carlo (MCMC) sampling \citep{2017A&A...600A..78P}. Although this fit is in a general agreement with the observed signal above 1.69\,$R_\odot$ too (see the top panel of Fig.~\ref{fig:wavelet_1}), it is indistinguishable from the noise dominating at this region. The bottom panel of Fig.~\ref{fig:wavelet_2} shows the shorter-wavelength periodic component filtered out from the wavelet spectrum in comparison with the longer-wavelength one obtained from fitting. The longer-wavelength variation clearly correlate with the envelope of the shorter-wavelength variation. It should be also mentioned here that the imaging information (independent of the density model of the atmosphere) cannot be used for an alternative measurement of the distances between the wave peaks in the discussed event, as the obtained wavelengths (about $0.003\,R_\odot\approx 2$\,Mm and $0.017\,R_\odot\approx 12$\,Mm) are far too short to be resolved with LOFAR. 

\begin{figure}
	\begin{center}
		\includegraphics[width=\linewidth]{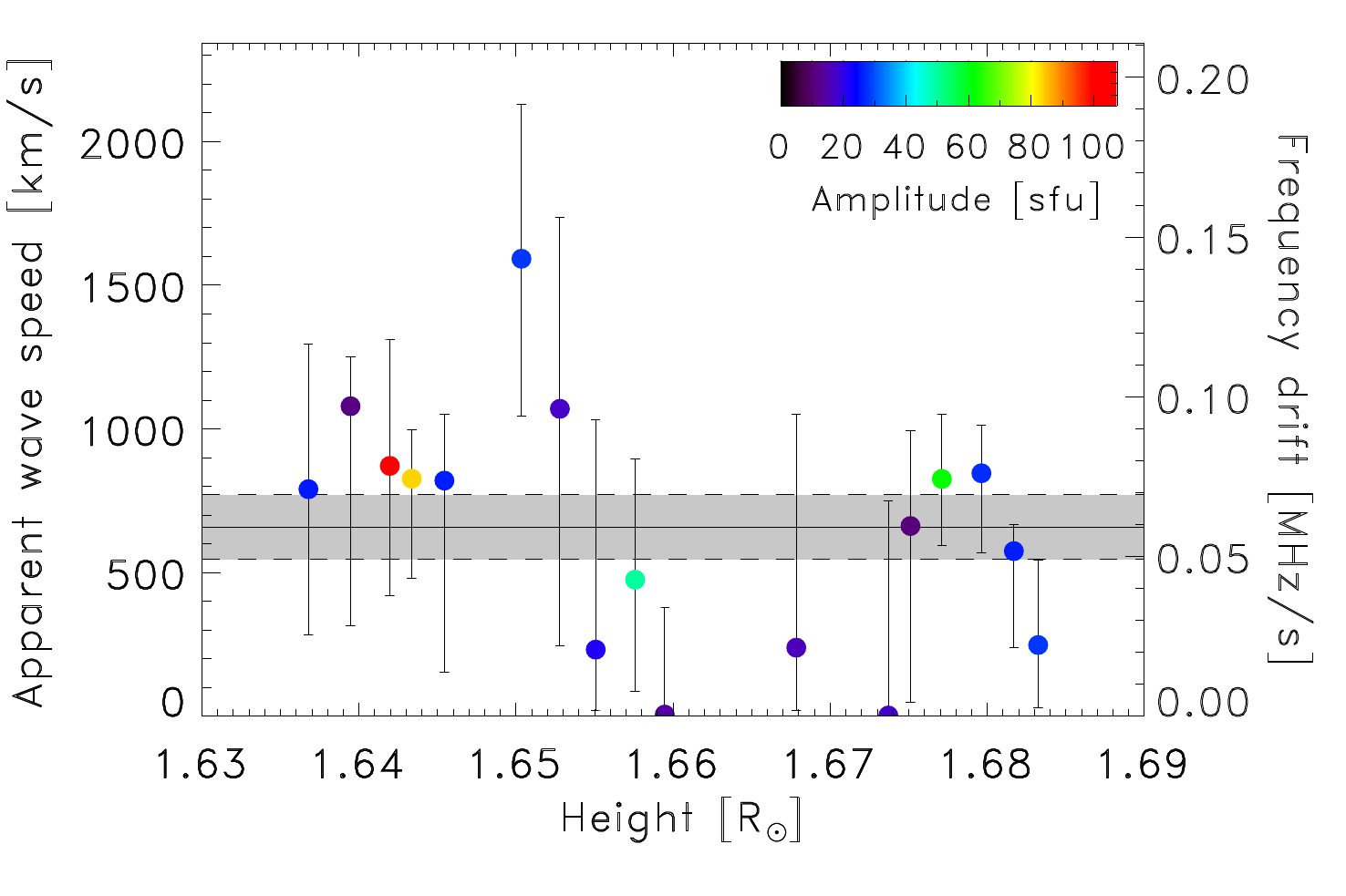}
	\end{center}
	\caption{Frequency drifts and apparent wave speeds estimated from the slopes of each individual stria highlighted in the right-hand panel of Fig.~\ref{fig:spectra}, and their mean values of $0.06\pm0.01$~MHz\,s$^{-1}$ and $657\pm114$~km\,s$^{-1}$, respectively, with the associated uncertainties shown by the shaded area. The colour scheme corresponds to the flux amplitude in each individual stria.
	}
\label{fig:speed}
\end{figure}

Using the dynamic spectrum shown in Fig.~\ref{fig:spectra}, one can estimate the drift rate $df/dt$ of each individual stria (which is apparently present and non-zero) in the region of interest, namely from 1.63\,$R_\odot$ to 1.69\,$R_\odot$, where the quasi-periodic behaviour is the most pronounced. Furthermore, using the dependence of the plasma frequency upon the density given by Eq.~(\ref{eq:freq}) and the Newkirk model of the solar atmosphere given by Eq.~(\ref{eq:newkirk}), one can estimate the phase speed $v$ at which these fluctuations propagate outwards from the Sun, $v=(2\Lambda/f)(df/dt)$, where $\Lambda\approx 0.3R_\odot$ is the characteristic density scale height \citep[e.g.,][]{2017NatCo...8.1515K}. We apply this approach to 17 the brightest individual striae (i.e. those with the radio flux above 20\% of the maximum flux in the region of interest), and estimate an apparent phase speed for each of them (see Fig.~\ref{fig:speed}). The estimation shows a rather scattered behaviour of the speed around its mean value which is approximately $657\pm114$~km\,s$^{-1}$, with no systematic variation of it with height in the considered interval. Moreover, Fig.~\ref{fig:speed} shows no dependence of the apparent speed upon the flux amplitude in each stria, suggesting the propagating wave is of a linear nature.

Finally, having obtained the wavelength of the propagating, shorter-period component, $(3\pm0.8)\times10^{-3}$\,$R_\odot$, and an apparent propagation speed, $657\pm114$~km\,s$^{-1}$, we can calculate its oscillation period as $2.9\pm 1.0$\,s, and conclude on the wave type, which is discussed in the next Section. 

\section{Discussion of observational results} \label{sec:dics}

\subsection{Physical mechanism} \label{sec:mech}

We begin with the discussion of the shorter-wavelength variation of the radio emission. The estimated properties of this oscillation (see Sec.~\ref{sec:obs}), such as its wavelength, $\sim2$\,Mm, propagation speed, $\sim657$\,km\,s$^{-1}$, and oscillation period, $\sim3$\,s, suggest it to be in the realm of the MHD description. In this context, taking into account that in the corona the plasma-$\beta$ is usually smaller than unity, the shorter-period oscillatory pattern is likely to be associated with the fast band MHD waves. MHD waves that propagate at the speed about the Alfv\'en speed are fast magnetoacoustic or Alfv\'en waves. We consider these wave modes as potential candidates for the observed oscillation. More specifically, the observed modulation of the radio emission can be caused either by the variation of the local magnetic field orientation \cite[e.g. by Alfv\'en waves, see e.g.][]{2014A&A...562A..57R} or by the variations of the plasma density {\citep{1972A&A....20...55D}} via fast waves, or by their combination.

In the first option, the magnetic field does not directly affect the intensity of the Langmuir waves excited by the bump-on-tail instability, but it may play a role in their conversion into the electromagnetic waves. However, the observed coherency of the oscillation suggests to exclude Alfv\'en waves, which are essentially non-coherent due to a local nature and phase mixing.
{Indeed, Alfv\'en waves propagating on neighbouring magnetic surfaces are decoupled, and hence amplitudes, spectra and phases of these waves do not correlate with each other \citep[see e.g.][]{2016SSRv..200...75N}.}
It is also unclear how to explain the observed periodicity if it is produced by Alfv\'en waves, as there have not been detections of coronal or chromospheric Alfv\'en waves with the detected periodicity {\citep[see e.g.][]{2015SSRv..190..103J,2016GMS...216..431V}}.

On the other hand, fast magnetoacoustic waves guided by plasma non-uniformities are collective compressive motions of plasma, which readily modulate Langmuir waves via the density perturbations. The detected speed is consistent with the previous observations of propagating fast waves in the corona in the EUV band \citep[cf. $600$\,km\,s$^{-1}$ to $735$\,km\,s$^{-1}$ estimated by][]{2013A&A...554A.144Y}. Likewise, the detected periodicity can be naturally produced by the dispersive evolution of fast waves in a magnetoacoustic waveguide, eventually forming fast wave trains. The detected period is also consistent with the theoretical estimations \citep[see, e.g.,][for recent theoretical results, and references therein]{2016ApJ...833...51Y, 2017ApJ...836....1Y, 2018ApJ...855...53L} and observations of these waves in the visible light \citep[cf. 6\,s detected by][]{2002MNRAS.336..747W} and microwaves \citep[cf. 2\,s detected by][]{2011SoPh..273..393M}. Moreover, a fast wave could easily modulate the radio emission by the variations of the local electron plasma frequency.  For example, the harmonic density wave with a relative amplitude of 0.1\% can cause almost 50\% modulations in Langmuir waves, as was shown by numerical simulations \citep[see, e.g., Fig.~6 of ][]{2001A&A...375..629K}. Thus, even small-amplitude density fluctuations can cause striae in the spectrum of electromagnetic radiation created by Langmuir waves in a structured atmosphere. 
\begin{figure}
	\begin{center}
		\includegraphics[width=\linewidth]{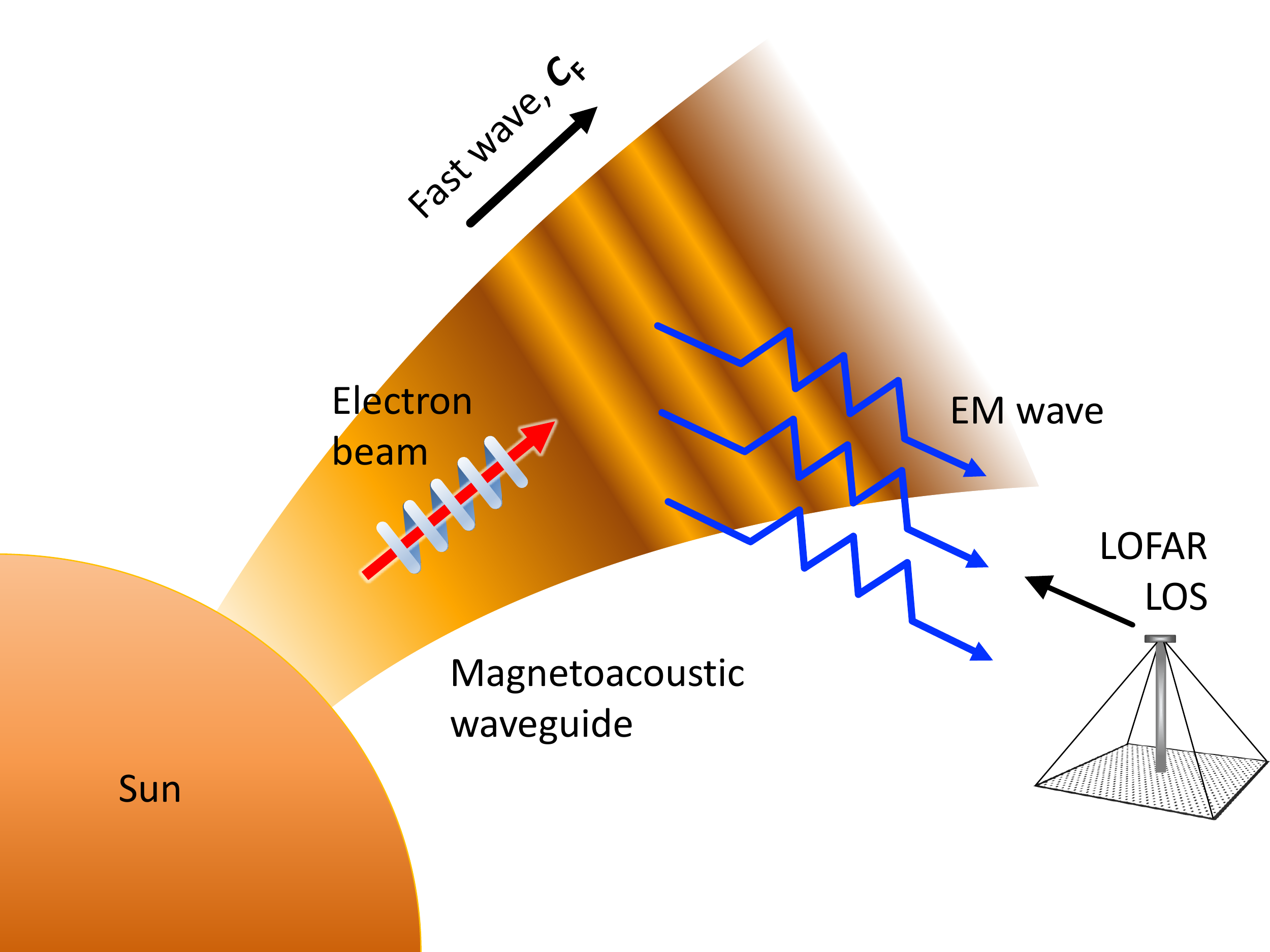}
	\end{center}
	\caption{A schematic synopsis illustrating a qualitative scenario of the generation of quasi-periodic striation in a dynamic spectrum of the type III burst by a propagating fast magnetoacoustic wave train. In this scenario, a broadband fast magnetoacoustic wave propagates along a field-aligned magnetic non-uniformity, acting on it as a waveguide, and gradually evolves in a quasi-periodic wave train due to the waveguide dispersion. An electron beam follows the same magnetic flux tube, and interacts with the plasma. The plasma concentration is locally modulated by the fast wave train.  The beam-plasma interaction generates the quasi-periodically modulated radio emission, observed by LOFAR.}
	\label{fig:sketch}
\end{figure}

A schematic sketch qualitatively illustrating the discussed scenario is presented in Fig.~\ref{fig:sketch}. It shows a fast magnetoacoustic wave guided by a plasma non-uniformity stretched along an open magnetic field, acting as a fast magnetoacoustic waveguide. Due to the presence of a characteristic spatial scale, i.e. the width of the waveguide, a broadband fast wave experiences the dispersive evolution, and eventually forms a quasi-periodic wave train \citep[see e.g.][for numerical simulations and observations of this effect in the corona]{2005SSRv..121..115N, 2013A&A...560A..97P, 2014A&A...569A..12N}. A beam of non-thermal electrons propagates upwards along an open magnetic field, through the ambient plasma density that is locally modified quasi-periodically by this fast wave train. The interaction of the non-thermal electron beam with the background plasma generates Langmuir waves with the frequency equal to the local electron plasma frequency. The Langmuir waves are converted into the electromagnetic waves received by LOFAR. Thus, the quasi-periodic spatial pattern of the modulated plasma density, and hence of the local electron plasma frequency, created by the fast wave train in a certain range of coronal heights appears in the dynamic spectrum of the observed radio emission. 

\begin{figure}
	\begin{center}
		\includegraphics[width=\linewidth]{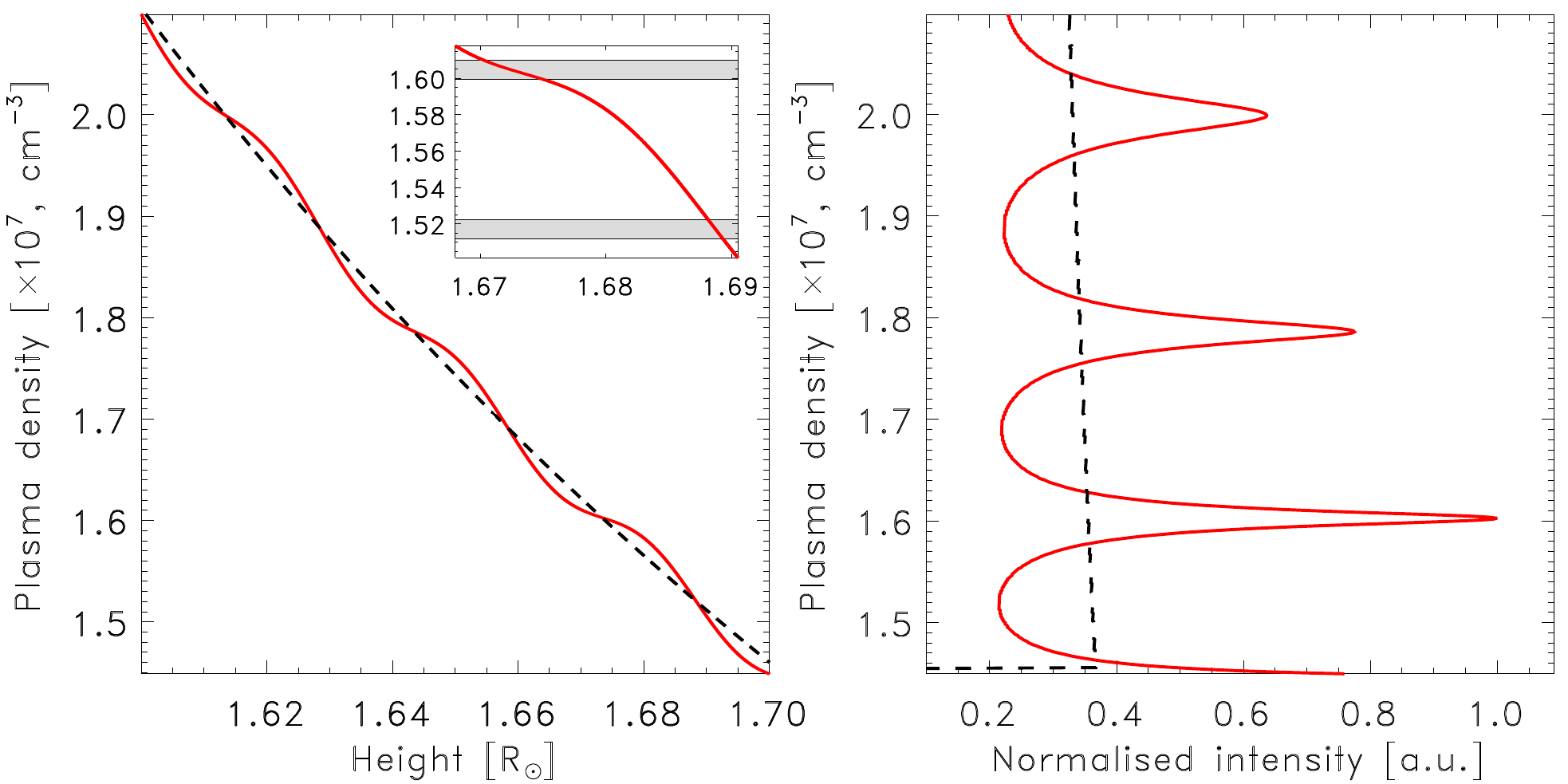}
	\end{center}
	\caption{Mechanism for the generation of quasi-periodic striae in the observed dynamic spectrum of a type III radio burst, based on the redistribution of the radio emission intensity on spatially quasi-periodic plasma density inhomogeneities. The calculations are based on the assumption that the emission intensity in a certain 12~kHz-wide frequency channel is proportional to the amount of plasma in the emitting volume with the electron plasma frequencies in the band of the channel. The shaded areas show the LOFAR spectral resolution, 12~kHz-wide frequency channels (multiplied by a factor of 10 in this figure for a better visualisation), within which the emission intensity is calculated according to Eq.~(\ref{eq:emissionmodel}). The black dashed lines show the unperturbed Newkirk plasma density profile (left) and the corresponding emission intensity (right). The red solid lines show the Newkirk plasma density perturbed by a harmonic density oscillation with the relative amplitude of 1\% and the wavelength of $0.03\,R_\odot$ (left), and the corresponding emission intensity (right), whose peaks occur in the regions of the lowest density gradient. Both lines in the right-hand panel are normalised to the maximum value of the red line.
	}
	\label{fig:model}
\end{figure}

Now we demonstrate how the mechanism described above could generate narrow, drifting quasi-periodic striae in the observed dynamic spectrum (Fig.~\ref{fig:spectra}). Assume that the observed striation is produced by the modulation of the background plasma density determined by the Newkirk model, see Eq.~(\ref{eq:newkirk}), by a quasi-monochromatic small-amplitude density perturbation. Given the fact that LOFAR operates in 12~kHz-wide frequency channels, and assuming the optically thin emission, we reckon that the observed emission intensity variation $\Delta I$ is proportional to the amount of plasma in the emitting volume, i.e. 
\begin{equation}
\label{eq:emissionmodel}
\Delta I \propto \Delta n \Delta r,
\end{equation}
where $\Delta n$ corresponds to the snapshot of the variation of plasma density in a 12~kHz-wide frequency channel, appearing at the vertical extent $\Delta r$. The value of $\Delta r$ is determined by the heights at which the local electron plasma frequency is in the range of the frequency channel. Thus, applying this assumption to an unperturbed plasma with the density given by the Newkirk model, we obtain a gradual monotonic variation of the observed emission intensity, $\Delta I$, with the observational frequency (see Fig.~\ref{fig:model}). If the Newkirk density profile is modulated by a harmonic density oscillation with a relative amplitude of 1\% and the wavelength of $0.03\,R_\odot$, the dynamic spectrum of the generated radio emission has a quasi-periodic behaviour of $\Delta I$ as a function of frequency. The increased spectral intensity occurs in the regions of the lowest density gradient as seen also in Fig.~\ref{fig:model}.

The longer-wavelength modulation of the shorter-wavelength quasi-monochromatic signals, i.e. the quasi-periodic clustering of the striae in three distinct groups as seen in the dynamic spectrum in the right-hand panel of Fig.~\ref{fig:spectra} may be caused by several different reasons. Unfortunately the phase speed of these longer-wavelength perturbations could not be estimated observationally, because of the insufficient height range of their detection. It does not allow us to use this information in the interpretation. Thus, we consider three options that could give the variation of the plasma density, required for the interpretation of the clustering of striae in the dynamic spectrum. One option is the effect of clumping of Langmuir waves due to stationary density fluctuations, akin to that observed in the solar wind \citep{1999JGR...10417069K}. However, such fluctuations have not been detected in the corona. Another option could be the presence of slow magnetoacoustic waves. Usually, those waves are seen at much lower heights \citep[see, e.g.][for a comprehensive review]{2009SSRv..149...65D}, but they have been also considered for the interpretation of the compressive perturbations detected at the coronagraph heights \citep[e.g.,][]{2000ApJ...529..592O}. In this case, taking the plasma temperature to be about 2~MK, we obtain the sound speed of about 215~km\,s$^{-1}$. Together with the wavelength of 12~Mm, it allows us to estimate the wave period as about 55~s. However, the detected periods of coronal slow magnetoacoustic waves are a few minutes or longer. Another option would be another fast wave, similar to the shorter wavelength one discussed above. But, in this case the coexistence of two periodicities should be explained somehow. Thus, all those three interpretations of the longer-period variability have certain shortcomings. In addition, the compressive flows that modulate the plasma density and hence the observed radio emission could be also produced by the ponderomotive force in a nonlinear torsional Alfv\'en waves \citep[e.g.][]{1998JGR...10323677O, 2017ApJ...840...64S}, which in principle can also produce the observed striation. However, the questions of whether this effect can produce the observed periodicity, and the consistency of the observed results with the intrinsic lack of a collective behaviour in Alfv\'en waves remain open.

\begin{deluxetable}{lccc}
	\tablecaption{{Parameters of the model (\ref{eq:emissionmodel})--(\ref{eq:twopert}), where the shorter-wavelength component in the density perturbation is considered to propagate at the observed speed of the wave, while the longer-wavelength one is assumed to be either stationary (case \#1), or propagating at the sound speed (case \#2), or at the observed speed of the shorter-wavelength component (case \#3). The frequency drift, $df/dt$, of striae in the simulated spectra is estimated in each of these cases.}
		\label{table: fit}
	}
	\tablewidth{0pt}
	\tablehead{
		\colhead{} &
		\colhead{Case \#1} &
		\colhead{Case \#2} &
		\colhead{Case \#3}
	}
	\startdata
	$V_1$, km\,s$^{-1}$& 0 &213 & 657\\
	$V_2$, km\,s$^{-1}$& 657 &657 & 657\\
	$\lambda_1$, $R_\odot$&0.017 &0.017  &0.017  \\
	$\lambda_2$, $R_\odot$&0.003 &0.003 & 0.003\\
	$A_1$, \% & $5.1_{-0.7}^{+0.4}$ &$5.1_{-0.6}^{+0.7}$ & $5.1_{-0.5}^{+0.7}$\\
	$A_2$, \% & $0.34_{-0.09}^{+0.07}$ &$0.35_{-0.10}^{+0.07}$ & $0.36_{-0.11}^{+0.09}$ \\
	$df/dt$, MHz\,s$^{-1}$& 0.03--0.04 & 0.04--0.05 & 0.06--0.07
	\enddata
\end{deluxetable}

We now consider a more sophisticated perturbation of the plasma density, by a superposition of two harmonic compressive propagating disturbances. We take into account displacements of those disturbances in time at certain phase speeds, aiming to reproduce the observed dynamic spectrum discussed in Sec.~\ref{sec:obs}. Consider the radio emission from a background Newkirk profile of the plasma, perturbed by two harmonic waves with different amplitudes and wavelengths, and propagating at different phase speeds,
\begin{align}
\label{eq:twopert}
&\delta{n}(r,t)\propto A_1\cos\left[\frac{2\pi}{\lambda_1}(r-V_1t)+\phi_1\right]+\\
& A_2\cos\left[\frac{2\pi}{\lambda_2}(r-V_2t)+\phi_2\right],\nonumber
\end{align}
whose parameters are summarised in Table~\ref{table: fit}. We considered three different cases, where the longer-wavelength component, with the wavelength $\lambda_1$, is assumed to be either stationary, or propagating at the sound speed (see Sec.~\ref{sec:est} for its estimation), or at the observed speed of the shorter-wavelength disturbance. The shorter-wavelength component, with the wavelength $\lambda_2$, is considered to propagate at the phase speed determined by the drift of the spectral striae in Sec.~\ref{sec:obs}.

\begin{figure}
	\begin{center}
		\includegraphics[width=\linewidth]{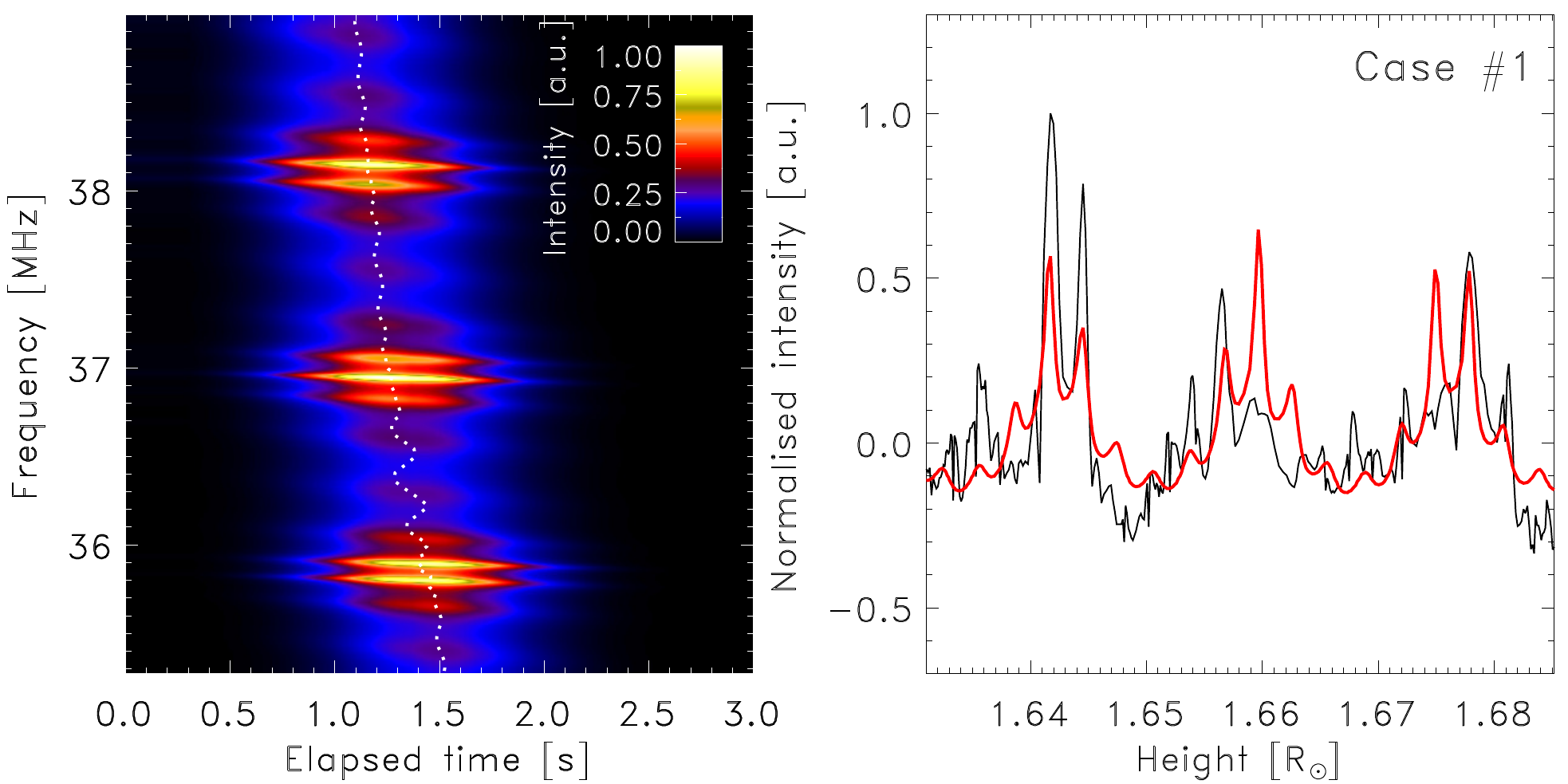}
		\includegraphics[width=\linewidth]{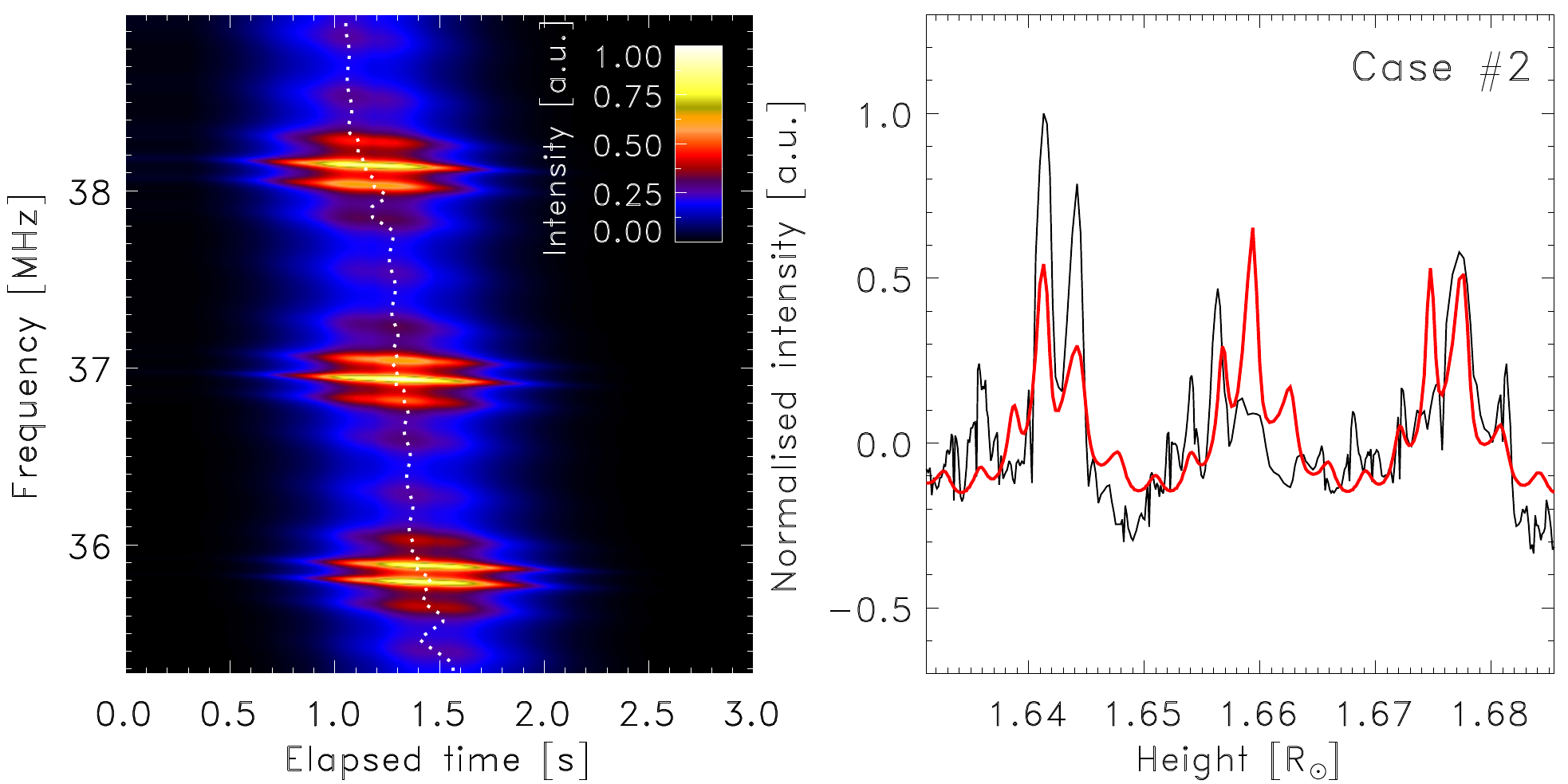}
		\includegraphics[width=\linewidth]{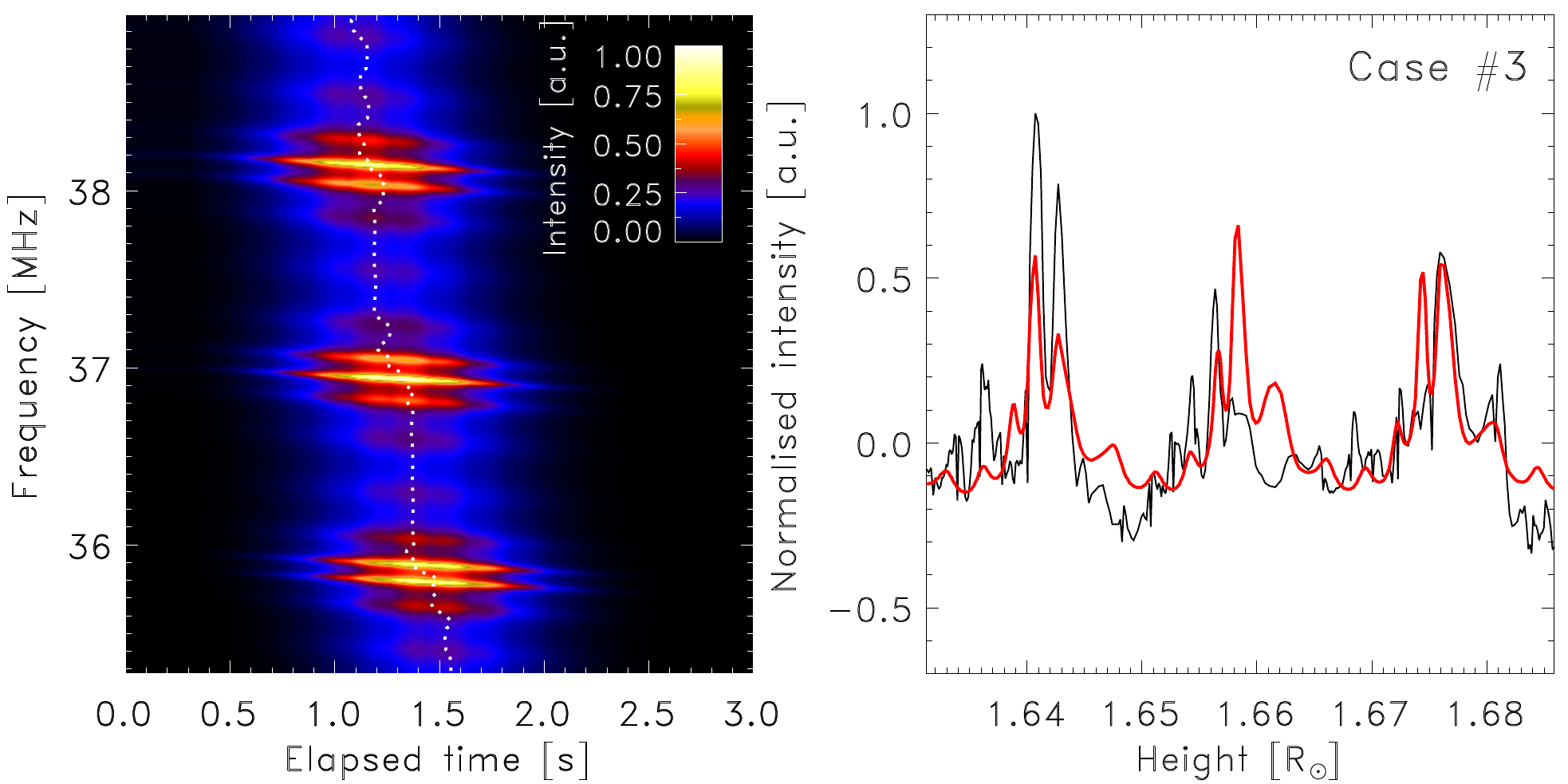}
	\end{center}
	\caption{Left: Modelled dynamic spectra produced by the mechanism shown in Fig.~\ref{fig:model} and mimicking that shown in Fig.~\ref{fig:spectra}, where the perturbation of the plasma density is taken in the form of Eq.~(\ref{eq:twopert}), with the parameters summarised in Table~\ref{table: fit} and corresponding to three different cases.
	The white dotted lines show spines of the modelled spectra, similar to that in Fig.~\ref{fig:spectra}.
	Right: A modelled emission intensity (red) calculated along the spines of the spectra shown in the left-hand panels in each of the three cases considered, in comparison to the observed radio flux (black) along the spine of the type III burst (see Figs.~\ref{fig:spectra}--\ref{fig:wavelet_2}), normalised to its maximum and with the overall trend subtracted.
	}
		\label{fig:modelspec}
\end{figure}

Fitting of model (\ref{eq:emissionmodel}) with the density perturbation in form (\ref{eq:twopert}) into the observed radio spectrum is shown in the right-hand panels of Fig.~\ref{fig:modelspec}, revealing a reasonable agreement between the modelling results and observational data for all three cases considered. For fitting, we use Bayesian inference and MCMC sampling technique referred to in Sec.~\ref{sec:obs}, which allows us to adequately estimate the relative amplitudes, $A_1$ and $A_2$, and constrain their uncertainties within 95\% credible intervals. For all three cases considered, the amplitudes are found to be stable within the estimated uncertainties (see Table~\ref{table: fit}), and we therefore stick on their mean values with the largest available uncertainties, i.e. $A_1=5.1_{-0.7}^{+0.7}\%$ and $A_2=0.35_{-0.11}^{+0.09}\%$, in the following discussion.
The dynamic spectra that are the modelled emission intensity in the frequency--time plane, in which this fit corresponds to the emission along its spine, is shown in the left-hand panels of Fig.~\ref{fig:modelspec} for each of these three cases. It is clearly seen to contain slightly inclined striation produced by the effect of shorter-wavelength waves, which is similar to that observed in Fig.~\ref{fig:spectra}. Moreover, the obtained striae are seen to cluster in three distinct groups, which is the effect of the superposition of two harmonic density perturbations with distinctly different wavelengths and amplitudes, as described above.
The frequency drift, $df/dt$, of the striae clusters is seen to increase with the propagation speed of the longer-wavelength component (see Table~\ref{table: fit}), being comparable to the observed value (0.05--0.07\,MHz\,s$^{-1}$, see Fig.~\ref{fig:speed}) in the cases \#2 and \#3.

A similar mechanism based on the redistribution of the plasma emission intensity on plasma density inhomogeneities was proposed in the application to the decimetric type IV bursts, producing so-called intermediate drift (or fiber) bursts \citep[cf. Fig~7 and Fig.~5 in][respectively]{1990A&A...236..242T, 2006SoPh..237..153K}. More specifically, super-Alfv\'enic solitons \citep{1990A&A...236..242T} and small-scale travelling MHD waves \citep{2006SoPh..237..153K} were considered as potential agents for producing quasi-periodic fine structures superimposed on the broadband type IV bursts, while in this paper we employ propagating fast MHD waves for the interpretation of the observed quasi-periodic striation in the metric type III burst.

\subsection{Estimation of the wave energy} \label{sec:est}

\begin{figure*}
	\begin{center}
		\includegraphics[width=0.8\linewidth]{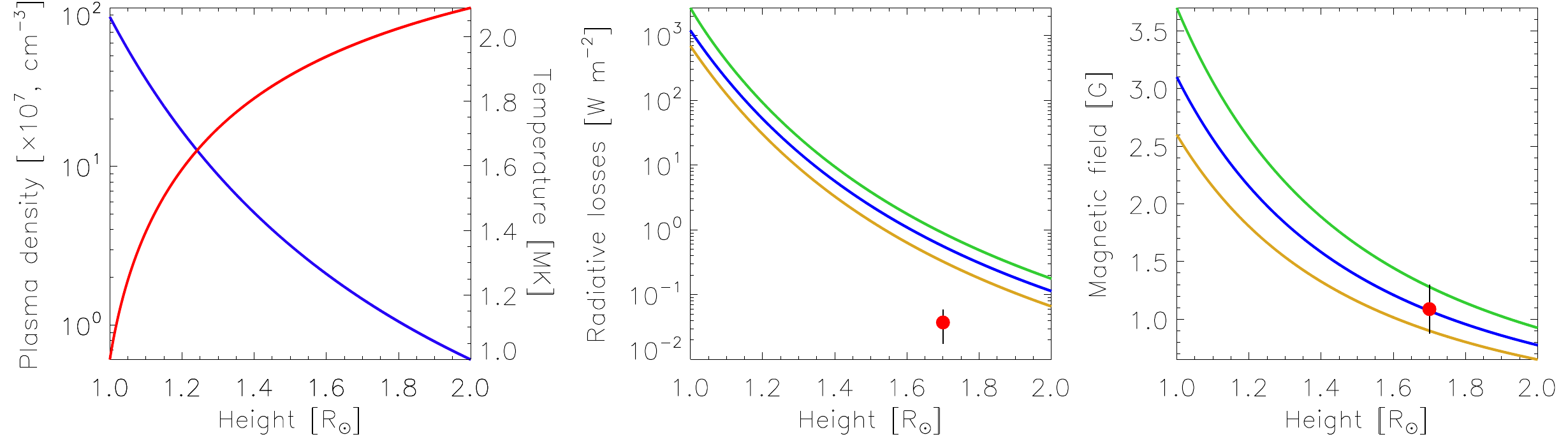}
	\end{center}
	\caption{Left: Model of the solar atmosphere, i.e. variation of density (the blue line) and temperature (the red line) with the heliocentric distance. The density variation is governed by the Newkirk model, the temperature is determined by Eq.~(\ref{eq:temp}) with $T_0=1$\,MK, $F_0=100$\,W\,m$^{-2}$, and $\kappa_0=10^{-11}$\,W\,m$^{-1}$\,K$^{-7/2}$ \citep{1996ApJ...461L.115S,1997ApJ...482..510W}.
	Middle: the radiative losses determined by Eq.~(\ref{eq:losses}), varying with height in the solar atmosphere, for three different temperatures of the base of the corona, $T_0=1$\,MK (green), 5\,MK (blue), and 15\,MK (yellow). For $T_0=1$\,MK, the lower corona ($r=1.2\,R_\odot$) is characterised by the losses of about 100\,W\,m$^{-2}$ \citep[cf.][]{1977ARA&A..15..363W}, while at $r=1.7\,R_\odot$ it drops to  about 1\,W\,m$^{-2}$. The red circle shows the estimated energy flux of the observed propagating fast magnetoacoustic wave, where the uncertainties are associated with those in the estimation of the wave amplitude (see Sec.~\ref{sec:mech}).
	Right: Estimation of the magnetic field from Eq.~(\ref{eq:alfvspeed}){, indicated by the red circle}. Black vertical lines show the magnetic field uncertainties associated with those in the wave speed estimation (see Fig.~\ref{fig:speed}). The curved lines show the radial model of the magnetic field, $B(r)=B_0(R_\odot/r)^2$, with $B_0=3.7$\,G (green), 3.1\,G (blue), and 2.6\,G (yellow).
	}
	\label{fig:energy_mf}
\end{figure*}

In this section we perform estimations of the fast MHD wave energy flux, based on the mechanism described in Sec.~\ref{sec:mech}, and compare it with the radiative energy losses. For that, we consider the model of the solar atmosphere at heliocentric {distances} of 1--2\,$R_\odot$, shown in the left-hand panel of Fig.~\ref{fig:energy_mf}. In this model, the Newkirk profile given by Eq.~(\ref{eq:newkirk}) is used for the variation of plasma density with height, while the coronal temperature variation is taken from \citet{1996ApJ...461L.115S} and \citet{1997ApJ...482..510W}, so that 
\begin{equation}
\label{eq:temp}
T(r)=\left[T_0^{7/2}+\frac{7R_\odot F_0}{2\kappa_0}\left(1-\frac{R_\odot}{r}\right)\right]^{2/7},
\end{equation}
where $T_0$ and $F_0$ are the temperature and the inward heat flux at the base of the corona, $r=R_\odot$, and $\kappa_0$ is the coefficient of thermal conductivity taken to be constant. In this assumption the temperature grows from 1\,MK at $r=R_\odot$ to about 2\,MK at $r=1.7R_\odot$ as seen in the left-hand panel of Fig.~\ref{fig:energy_mf}. Thus, we fix $r=1.7R_\odot$ as an approximate frontier of the region of interest in which the discussed quasi-periodic structure is the most pronounced (see Figs.~\ref{fig:flux}--\ref{fig:wavelet_2}). Table~\ref{table} shows the values of coronal parameters at this height, used for the following estimations.

\begin{deluxetable}{ll}
	\tablecaption{Typical values of the coronal plasma parameters at the heliocentric distance $r=1.7R_\odot$, obtained with the use of Eqs.~(\ref{eq:freq}), (\ref{eq:newkirk}), (\ref{eq:temp}), and  (\ref{eq:losses}), where $B$ denotes the value of the magnetic field measured in gauss.
		\label{table}
	}
	\tablewidth{0pt}
	\tablehead{
		\colhead{Parameter} &
		\colhead{Value}
	}
	\startdata
	Height, $r$ & 1.7$R_\odot$  \\
	Plasma density, $\rho$ & $2.4\times 10^{-14}$\,kg\,m$^{-3}$\\
	Plasma frequency, $f_\mathrm{pe}$ & 34\,MHz\\
	Temperature, $T$ & 2\,MK\\
	Adiabatic index, $\gamma$ & 5/3\\
	Sound speed, $C_\mathrm{S}$ & 213\,km\,s$^{-1}$\\
	Alfv\'en speed, $C_\mathrm{A}$ & $B$[G]$\times$571\,km\,s$^{-1}$\\
	Radiative energy losses, $\mathcal{Q}$ & 1\,W\,m$^{-2}$
	\enddata
\end{deluxetable}

We also adapt the following model for the local radiative energy losses per unit area (i.e. the radiative energy flux measured in W\,m$^{-2}$) at a certain height $r$ from the centre of the Sun,
\begin{equation}
\label{eq:losses}
\mathcal{Q}(r)=\mathcal{L}(n,T)\,r,
\end{equation}
in which the energy loss rate per unit volume, $\mathcal{L}$~(W\,m$^{-3}$), is taken in the form $\mathcal{L}(n,T)=\alpha n^2/\sqrt{T}$ \citep{2016ApJ...833...76B}, where $n$ and $T$ vary with height by Eqs.~(\ref{eq:newkirk}) and (\ref{eq:temp}), respectively, and a constant coefficient $\alpha=5\times10^{-31}$\,W\,m$^{2}$\,K$^{1/2}$ is chosen to satisfy the condition $\mathcal{Q}\approx100$\,W\,m$^{-2}$ at $r=1.2\,R_\odot$ (i.e. in the lower corona, $r<1.5\,R_\odot$) with the temperature at its base $T_0=1$\,MK \citep[see Table~1 in][]{1977ARA&A..15..363W}. The behaviour of the radiative energy flux $\mathcal{Q}$ determined by Eq.~(\ref{eq:losses}) with height in the solar atmosphere is illustrated in the middle panel of Fig.~\ref{fig:energy_mf} for different values of the base temperature $T_0$. It is a decreasing function varying from about 100\,W\,m$^{-2}$ at $r=1.2\,R_\odot$ to 1\,W\,m$^{-2}$ at $r=1.7\,R_\odot$ for $T_0=1$\,MK.

Assuming that the observed wave is a fast magnetoacoustic wave propagating at the phase speed $657$\,km\,s$^{-1}$ (see Secs.~\ref{sec:obs} and \ref{sec:mech}), which is about the local fast speed $C_\mathrm{F} = \sqrt{C_\mathrm{A}^2 +C_\mathrm{S}^2}$, we can deduce the characteristic Alfv\'en speed $C_\mathrm{A}$ at the observed height (see Table~\ref{table}). We obtain $C_\mathrm{A}=\sqrt{C_\mathrm{F}^2-C_\mathrm{S}^2}\approx622$\,km\,s$^{-1}$. Furthermore, taking into account the observed relative wave amplitude $\delta n/n\approx0.35$\% (see Sec.~\ref{sec:mech}) and the fact that in a fast wave it is about the relative amplitude of the transverse velocity, $\delta V/C_\mathrm{A}$, which readily follows from the continuity condition, we obtain $\delta V \approx 2$\,km\,s$^{-1}$. Having the velocity amplitude calculated, we can estimate the propagating wave energy flux as
\begin{equation}
\label{eq:waveflux}
\mathcal{F} = \frac{1}{2}\rho\,\delta V^2C_\mathrm{F},
\end{equation}
where $\rho$ is the mass density at the observed height (see Table~\ref{table}). According to this estimation, $\mathcal{F} \approx 0.04\pm0.02$\,W\,m$^{-2}$ where the uncertainties are associated with those in the wave amplitude estimation (see Sec.~\ref{sec:mech}). The obtained value of $\mathcal{F}$ is at least an order of magnitude less than the radiative energy losses at this height (1\,W\,m$^{-2}$).

\subsection{Estimation of the magnetic field} \label{sec:est_mf}

The association of the observed effect with a fast magnetoacoustic wave allow us to estimate the magnetic field $B$ from the value of the Alfv\'en speed $C_\mathrm{A}$ deduced above. Given the definition of $C_\mathrm{A}$ and estimating the density by the electron plasma frequency, it reduces to
\begin{equation}
\label{eq:alfvspeed}
C_\mathrm{A}=\frac{B}{\sqrt{\mu_0\rho}}\approx B\mathrm{[G]}\times571\,\mathrm{km\,s}^{-1},
\end{equation} 
where $\mu_0=4\pi\times 10^{-7}$\,H\,m$^{-1}$ is the permeability of the vacuum. Thus, for $C_\mathrm{A}\approx622$\,km\,s$^{-1}$, we obtain $B\approx 1.1\pm0.2$\,G where the uncertainties are associated with those in the wave speed estimation (see {the right-hand panel in Fig.~\ref{fig:energy_mf}}). Comparing this result with the radial magnetic field model, $B(r)=B_0(R_\odot/r)^2$, and using the heliocentric height estimation by the Newkirk model, see  Table~\ref{table}, we obtain a good agreement of our estimation with the values of $B_0$ ranging from 2.6\,G to 3.7\,G with the mean value of 3.1\,G.

\section{Summary and conclusions} \label{sec:conc}

We considered a quasi-periodic striation in the dynamic spectrum of a metric type III solar radio burst, observed with LOFAR. The quasi-periodic pattern is detected between approximately 35\,MHz and 39\,MHz, which correspond to the range of heliocentric distances from 1.6\,$R_\odot$ to 1.7\,$R_\odot$, assuming the Newkirk density model of the solar atmosphere. By the detected dynamic properties of this oscillation, we suggest to associate it with a fast magnetoacoustic wave train guided by a plasma non-uniformity along an open magnetic field outwards from the Sun. Our findings can be summarised as follows:

\begin{itemize}
	
\item  Spectral analysis of the detrended dependence of the observed radio flux upon height shows the presence of a systematic oscillatory behaviour between approximately 1.63\,$R_\odot$  and 1.69\,$R_\odot$. The detected wavelengths are about 0.003\,$R_\odot$ ($\approx 2$\,Mm) and 0.017\,$R_\odot$ ($\approx 12$\,Mm). The latter was obtained by the fitting of a harmonic function into the observational signal. Above 1.69\,$R_\odot$, the radio flux behaves rather stochastically, with no pronounced periodic component.

\item The observed striae are slightly inclined compared to the slope of the burst itself, indicating the presence of a travelling wave with the phase speed much lower than the speed of the emitting electron beam. From the frequency drift of each individual stria in the region of interest (1.63--1.69\,$R_\odot$), we estimated the speed with which this wave propagates outwards from the Sun, whose mean value is about 657\,km\,s$^{-1}$.

\item Having the wavelength ($\approx 2$\,Mm) and the propagation speed (657\,km\,s$^{-1}$) of the wave estimated, we calculated its oscillation period as about 3\,s, {which coincides with the theoretical estimations and previous observations of fast magnetoacoustic wave trains in the visible light {\citep[e.g.][]{2002MNRAS.336..747W}}, and decimetric  and microwave radio emission {\citep[e.g.][]{2011SoPh..273..393M}} at much lower heights.}

\item The detected characteristic of this travelling wave suggest to associate it with one of the fast MHD modes. The Alfv\'en wave is very unlikely to produce the observed coherent oscillation due to its local nature and phase mixing. Moreover, it is unclear how the observed period can be explained by Alfv\'en waves. In contrast, the observed characteristics are consistent with the properties of dispersive fast magnetoacoustic wave trains propagating along a magnetic field non-uniformity, already detected in the solar corona.

\item We could not reveal the nature of the longer-wavelength component with the wavelength of about 0.017\,$R_\odot$ {($\approx 12$\,Mm)}. It can possibly be produced by the effect of clumping of Langmuir waves on density non-uniformities due to a slow magnetoacoustic wave, or another fast wave, while all those interpretations have certain shortcomings. In the case of the slow wave, its period would be about 55~s. 

\item We suggested a simple quantitative model explaining the observed modulation of the dynamic spectrum of the radio emission, based on the plasma emission mechanism, in which the electromagnetic emission intensity is assumed to be proportional to the amount of plasma in the emitting volume. The background plasma density perturbed by a wave {leads to the appearance of} spectral peaks in the observed intensity, which correspond to the emission coming from the regions of the lowest density gradient.

\item Fitting this model into the observed dynamic spectrum, we obtained the relative amplitude of the propagating fast wave train, which is about 0.35\% or 2\,km\,s$^{-1}$. 

\item We used this value of the wave amplitude and the detected wave speed to estimate the wave energy flux, which was found to be about 0.04\,W\,m$^{-2}$, and compared it to the local radiative energy losses, which are about 1\,W\,m$^{-2}$ at the considered heliocentric height 1.7\,$R_\odot$. Accounting for the estimation uncertainties, the obtained value of the wave energy flux is at least an order of magnitude less than the local radiative losses.

\item Treating the detected propagation speed of the wave as a fast speed and fixing other parameters of the plasma to their typical values at the observed height 1.7\,$R_\odot$, we deduced the value of the Alfv\'en speed at this height to be about 622\,km\,s$^{-1}$. Using this value, we obtained the magnetic field strength to be about 1.1\,G, which was found to be consistent with a radial model of the magnetic field.

\end{itemize}



The wavelength of the detected fast waves is too short to allow for the use of the imaging spectroscopy with LOFAR for their study. However, the spatially nonresolving observations interpreted as longer-period fast waves in other events \citep[see e.g.][]{2016A&A...594A..96G,2017ApJ...844..149K} suggest that the imaging spectroscopy with LOFAR could be applied to the analysis of similar events.

\begin{acknowledgements}
This work is supported by the  STFC consolidated grant ST/P000320/1 (DYK, VMN). 
VMN acknowledges the support by the  BK21 plus program through the National Research Foundation funded by the Ministry of Education of Korea. EPK is supported by the STFC consolidated grant ST/P000533/1. We also acknowledge support from the International Space Science Institute for the team \lq\lq Low Frequency Imaging Spectroscopy with LOFAR -- New Look at Non-Thermal Processes in the Outer Corona\rq\rq\footnote{\url{http://www.issibern.ch/teams/lofar/}}.
This paper is based (in part) on data obtained from facilities of the International LOFAR Telescope (ILT)
under project code LC3-012. LOFAR \citep{2013A&A...556A...2V} is the Low Frequency Array designed
and constructed by ASTRON. It has observing, data processing, and data storage facilities in several countries,
that are owned by various parties (each with their own funding sources), and that are collectively operated
by the ILT foundation under a joint scientific policy. The ILT resources have benefitted
from the following recent major funding sources: CNRS--INSU, Observatoire de Paris
and Université d'Orléans, France; BMBF, MIWF--NRW, MPG, Germany; Science Foundation Ireland (SFI),
Department of Business, Enterprise and Innovation (DBEI), Ireland; NWO, The Netherlands;
The Science and Technology Facilities Council, UK.
\end{acknowledgements}

\bibliographystyle{aasjournal} 



\end{document}